\documentstyle[12pt]{article}

\newcommand{\EQ}{\begin{equation}}
\newcommand{\EN}{\end{equation}}

\renewcommand{\thefootnote}{\fnsymbol{footnote}}
\def\aprle{\buildrel < \over {_{\sim}}}
\def\aprge{\buildrel > \over {_{\sim}}}
\begin{document}
\topmargin 0pt
\oddsidemargin=-0.4truecm
\evensidemargin=-0.4truecm
\newpage
\setcounter{page}{0}
\begin{titlepage}
\begin{flushright}
SISSA 170/92/EP\\
\end{flushright}
\vspace{-0.6cm}
\begin{center}
{\large NEUTRINOS WITH MIXING IN TWISTING MAGNETIC \\
              FIELDS}
\vspace{-0.4cm}
\end{center}
\flushright{\it In memoriam of Ya.A. Smorodinsky}\\
\begin{center}
{\large E.Kh. Akhmedov
\footnote{On leave from Kurchatov Institute of Atomic
Energy, Moscow 123182, Russia}
\footnote{E-mail: akhmedov@tsmi19.sissa.it, ~akhm@jbivn.kiae.su},
 ~S.T. Petcov
\footnote{Istituto Nazionale di Fisica Nucleare, Sezione di Trieste,
 Trieste, Italy}
\footnote{Permanent address: Institute of Nuclear Research and Nuclear
Energy, Bulgarian Academy of Sciences, BG-1784 Sofia, Bulgaria}\\}
{\em Scuola Internazionale Superiore di Studi Avanzati\\
Strada Costiera 11, I-34014 Trieste, Italy} \\
\vspace{0.4cm}
{\large and}\\
\vspace{0.4cm}
{\large A.Yu. Smirnov}\\
{\em Institute for Nuclear Research, Russian Academy of Sciences,
Moscow, Russia, and\\
International Centre for Theoretical Physics, I-34100 Trieste, Italy}\\
\end{center}

\begin{abstract}
Transitions in a system of neutrinos with vacuum
mixing and magnetic moments, propagating
in matter and transverse magnetic field, are considered. It is shown
that in the realistic case of magnetic field direction varying along the
neutrino path qualitatively new phenomena
become possible: permutation of neutrino
conversion resonances, appearance of resonances
in the neutrino-antineutrino
($\nu_{lL}\leftrightarrow\bar{\nu}_{lR}$) transition
channels, neutrino-antineutrino resonant conversion,
large amplitude $\nu_{lL}\leftrightarrow\bar{\nu}_{lR}$ oscillations,
merging of different resonances (triple resonances).
Possible phenomenological implications
of these effects are briefly discussed.
\end{abstract}
\vspace{1cm}
\vspace{.5cm}
\end{titlepage}
\renewcommand{\thefootnote}{\arabic{footnote}}
\setcounter{footnote}{0}
\newpage
\section{Introduction}
\indent Neutrinos with magnetic or transition magnetic moments ($\mu$)
propagating in transverse magnetic fields (${\bf B_{\bot}}$)
will undergo spin precession \cite{C,Shrock,VVO} or spin-flavour precession
\cite{ShV,VVO,AKHM1,LM}, respectively.
As is well known, both types of precession can have important
astrophysical implications. In the presence of matter,
the neutrino spin-flavour precession can be resonantly enhanced
\cite{AKHM1,LM}. The resonant spin-flavour precession (RSFP)
is analogous to
the MSW effect in neutrino oscillations \cite{MS,W}. If neutrino mixing
exists, an interplay between
the RSFP and the MSW effect is possible. In particular, the
matter-enhanced neutrino oscillations can assist the RSFP in giving rise
to time
variations of the solar neutrino flux
by improving the adiabaticity of the latter \cite{AKHM4}.

In most of the previous studies of the spin and spin-flavour precessions of
neutrinos in transverse magnetic fields
it was assumed that the direction of the field is fixed.
However, in the physically interesting cases of three-dimensional
field configurations (e.g., in the Sun, supernovae, etc.),
neutrinos traversing the magnetic field will feel, in general, a field whose
direction is changing along the neutrino trajectory \cite{VW,ASh1,Sm,AKS,
ASh2}. The presence of such twisting magnetic fields can lead to new
interesting phenomena. In
particular, the neutrino spin precession can be resonantly
enhanced, the energy dependence of the probability of the resonant
spin-flavour precession can be significantly changed, the lower limit
on the value of $\mu B_{\bot}$ required to account for the solar neutrino
problem can be slightly relaxed and, in addition, new realizations of RSFP
become possible \cite{Sm,AKS}.

The RSFP of neutrinos in twisting magnetic fields was considered so far
under the simplifying assumption of absence of
neutrino mixing in vacuum. In the present paper we study the general case:
we assume the existence not only of flavour-off-diagonal neutrino
magnetic moments, but also of flavour neutrino mixing
in vacuum. This is a natural possibility if no additive lepton number is
conserved. Under the above condition both resonant
oscillations and RSFP of
neutrinos are allowed. We show that if they take place in twisting
magnetic fields, qualitatively new
phenomena become possible: resonant enhancement of the
helicity-flipping $\Delta L_{l} = 2$ neutrino-antineutrino
conversions $\nu_{lL} \leftrightarrow \bar{\nu}_{lR}$
($l=e,\mu,\tau$), and merging of three
different transition resonances.
Moreover, for a definite sign of the angular velocity
of the magnetic field rotation, the order of the RSFP and MSW resonances
in matter with monotonically decreasing density can
change. All these effects can have an important impact on solar
and supernova neutrinos.

\section{Neutrino Evolution Equations}

\indent Consider a system of two flavour neutrinos
and their antiparticles,
say $\nu_{eL}$, $\bar{\nu}_{eR}$, $\nu_{\mu L}$, and
$\bar{\nu}_{\mu R}$.
We shall assume that $\nu_{e}$ -- $\nu_{\mu}$ mixing is generated by a
Majorana neutrino mass term, and that there exists a $\nu_{eL}$ --
$\bar{\nu}_{\mu R}$ transition magnetic moment $\mu$.
The equations for the flavour neutrino wave functions,
describing the propagation of neutrinos in matter and
transverse magnetic field in the case of interest, can be written in
the form:
  \EQ
i\frac{d}{dt}\left(\begin{array}{l}
   \nu_{eL}\\
   \bar{\nu}_{eR}\\
   \nu_{\mu L}\\
   \bar{\nu}_{\mu R}
\end{array}\right )
{}~=~\left (
\begin{array}{cccc}
   N(t)-c_{2}\delta & 0 & s_{2}\delta & \mu B_{\bot}(t)e^{i\phi (t)}\\
  0 & -N(t)-c_{2}\delta & -\mu B_{\bot}(t)e^{-i\phi (t)} & s_{2}\delta\\
  s_{2}\delta & -\mu B_{\bot}(t)e^{i\phi (t)} & -r(t)N(t)+c_{2}\delta & 0 \\
  \mu B_{\bot}(t)e^{-i\phi (t)} & s_{2}\delta & 0 & r(t)N(t)+c_{2}\delta
\end{array}\right )\left(\begin{array}{l}
   \nu_{eL}\\
   \bar{\nu}_{eR}\\
   \nu_{\mu L}\\
   \bar{\nu}_{\mu R}
\end{array}\right )
\EN
Here the angle $\phi (t)$ defines the direction of the magnetic field
${\bf B}_{\bot}(t)$ in the plane orthogonal to the neutrino momentum,
$B_{\bot}(t) = |{\bf B}_{\bot}(t)|$,
$$N\equiv \sqrt{2}G_{F}(n_{e}-n_{n}/2),~~r\equiv \frac {n_{n}}
{2n_{e}-n_{n}},~~\delta\equiv \Delta m^{2}/4E,$$
\EQ
s_{2}\equiv\sin 2\theta_{0},~~c_{2}\equiv \cos 2\theta_{0},
\EN
where $G_{F}$ is the Fermi constant, $n_{e}$ and $n_{n}$ are the electron
and neutron number densities, $E$ is the neutrino energy,
$\Delta m^{2}=m_{2}^{2}-
m_{1}^{2}$, $m_1$ and $m_2$ being the masses of two Majorana neutrinos
and $\theta_{0}$ being the neutrino
mixing angle in vacuum. The zeros in the effective Hamiltonian
in eq. (1) are related to the fact that diagonal magnetic moments
of Majorana neutrinos are precluded by $CPT$ invariance.

It is instructive to consider the evolution of the neutrino
system in the reference frame
which rotates with the same angular velocity as the transverse magnetic
field \cite{Sm}. The corresponding transformation of the neutrino
wave functions reads: $\nu = S\nu'$, where $\nu$
is the column vector of state of the neutrino system in eq. (1), and
$S = $
diag$(e^{i\frac{\phi}{2}},
e^{-i\frac{\phi}{2}},e^{i\frac{\phi}{2}},e^{-i\frac{\phi}{2}})$.
Since $S$ is a diagonal matrix,
the above transformation does not change
the probabilities of the neutrino transitions and therefore we will keep
the same notations for the neutrino wave functions in the new frame.
In the rotating frame the phase factors disappear from the
nondiagonal terms (see eq. (1)) and the
evolution equations take the form:
\EQ
i\frac{d}{dt}\left(\begin{array}{l}
   \nu_{eL}\\
   \bar{\nu}_{eR}\\
   \nu_{\mu L}\\
   \bar{\nu}_{\mu R}
\end{array}\right )
{}~=~\left (
\begin{array}{cccc}
   N-c_{2}\delta+\frac{\dot{\phi}}{2} & 0 & s_{2}\delta & \mu B_{\bot}\\
  0 & -N-c_{2}\delta-\frac{\dot{\phi}}{2} & -\mu B_{\bot} & s_{2}\delta\\
  s_{2}\delta & -\mu B_{\bot} & -rN+c_{2}\delta+\frac{\dot{\phi}}{2} & 0\\
  \mu B_{\bot} & s_{2}\delta & 0 & rN+c_{2}\delta-\frac{\dot{\phi}}{2}
\end{array}\right )\left(\begin{array}{l}
   \nu_{eL}\\
   \bar{\nu}_{eR}\\
   \nu_{\mu L}\\
   \bar{\nu}_{\mu R}
\end{array}\right)
\EN
where ${\dot {\phi}}\equiv d\phi/dt$.

The 4$\times$4 matrix on the right-hand side of the Schroedinger-like
equation (3) defines the time-dependent
effective Hamiltonian of the system,
$H$. Note that the effect of
rotation of the magnetic field amounts
to the appearance of additional terms
of the same absolute value in the diagonal elements of $H$.
These terms have the
same signs for the neutrinos
possessing the same helicity and opposite
signs for the neutrinos of opposite helicities.
The indicated features
have a
simple physical interpretation \cite{AKS}.
According to classical mechanics, a
system with angular momentum  {\bf s}  acquires an additional
energy $\Delta E =- ({\bf s}{\bf \Omega})$ upon a
transformation to a reference frame
rotating with an angular velocity
${\bf \Omega}$. In our case $\Omega = \dot {\phi}$,
$s = 1/2$, and the projections of ${\bf s}$ on
${\bf \Omega}$ are equal to $-1/2$ for the neutrinos,
and to $+1/2$ for the antineutrinos.

Let us consider the resonant properties
of the neutrino system under
study. The diagonal elements of the Hamiltonian in eq. (3)
define the energies of the "flavour levels", i.e.
of $\nu_{eL}$, $\bar{\nu}_{eR}$, $\nu_{\mu L}$,
and $\bar{\nu}_{\mu R}$, in the absence of vacuum mixing
and magnetic field; we shall denote them respectively as
$ H_{e}$, $H_{\bar{e}}$, $H_{\mu}$,
and $H_{\bar{\mu}}$. Resonances take place at
densities at which the splittings
between the eigenvalues of the effective
Hamiltonian $H$ with $s_{2}\delta \ne 0$ and
$\mu B_{\bot}\ne 0$ are minimal. The
resonance densities
can be approximately
found by equating pairs of
diagonal elements of $H$:
$H_{\alpha}(n_{e},n_{n})=H_{\beta}(n_{e},n_{n})$,
$\alpha \ne \beta, \alpha, \beta = e,\bar{e},\mu, \bar{\mu}$
(i.e. they correspond
approximately to the values of $N$ at which the flavour
levels cross).

With four neutrino states there are altogether
six resonance conditions,
and in what follows we will consider the nature
and the properties of the
corresponding resonances.

Two conditions, $H_{e} = H_{\mu}$ and
$H_{\bar {e}} = H_{\bar {\mu}}$, can
be written as:
\begin{eqnarray}
\sqrt{2} G_{F} n_{e} &=& \frac{\Delta m^{2}}{2E}\, \cos 2\theta_{0}
{}~~~(\nu_{eL}\leftrightarrow \nu_{\mu L}), \\
\sqrt{2} G_{F} n_{e} &=& -\frac{\Delta m^{2}}{2E}\, \cos 2\theta_{0}
{}~~~(\bar{\nu}_{eR}\leftrightarrow \bar{\nu}_{\mu R}).
\end{eqnarray}
They concern the transitions (given in parentheses)
between neutrinos
having the same helicity but different flavour, and thus correspond
to the MSW resonances. The condition for the neutrinos (antineutrinos)
can be satisfied for $\Delta m^{2}\cos 2\theta_0>0$ ($\Delta m^{2}
\cos 2\theta_0<0$).
Eqs. (4) and (5) do not depend
on $\dot {\phi}$ since the flavour neutrino states involved have the
same helicity and the magnetic field rotation influences
both states equally. Magnetic fields (twisting
or nontwisting) do not modify the MSW resonances.

The next two conditions,
$H_{e} = H_{\bar {\mu}}$ and
$H_{\bar {e}} = H_{\mu}$, have the form:
\begin{eqnarray}
\sqrt{2}G_{F}(n_{e}-n_{n}) &=& \frac{\Delta m^{2}}{2E}
          \, \cos 2\theta_{0} -
\dot{\phi}~~~(\nu_{eL}\leftrightarrow \bar{\nu}_{\mu R}), \\
\sqrt{2}G_{F}(n_{e}-n_{n}) &=& - \frac{\Delta m^{2}}{2E} \,
             \cos  2\theta_{0} -
\dot{\phi}~~~(\bar{\nu}_{eR}\leftrightarrow \nu_{\mu L}).
\end{eqnarray}
The neutrinos involved in the relevant transitions
have different
flavours, as well as opposite helicities. Thus, eqs. (6) and (7) define
the resonance conditions for the spin-flavour precession:
eq. (6) refers to the channel $\nu_{eL}\leftrightarrow\bar{\nu}_{\mu R}$,
while eq. (7) refers to the $CP$-conjugated one. Since
the neutrinos involved have different helicities, the magnetic
field rotation modifies
the resonance conditions: both eq. (6) and eq. (7) depend
on $\dot{\phi}$. In a medium with
$(n_{e} - n_{n})$ of a definite sign
(like the matter of the Sun) and $\dot{\phi} = 0$
only one of these two conditions can be satisfied
depending on the sign of $\Delta m^{2}\cos 2\theta_0$. In contrast,
in the case of twisting
magnetic field both conditions can be obeyed
simultaneously. This can happen for $\dot{\phi}<0$
($\dot{\phi}>0$) when $n_{e} > n_{n}$
($n_{e} < n_{n}$) provided
$|\dot{\phi}| > |(\Delta m^{2}/2E) \cos 2\theta_{0}|$.
Obviously, the two conditions will
be fulfilled at different densities.

Finally, equating the elements $H_{e} = H_{\bar{e}}$
and $H_{\mu} = H_{\bar{\mu}}$ one finds:
\begin{eqnarray}
\sqrt{2}G_{F}(n_{e}-n_{n}/2) &=& -\dot{\phi}/2
{}~~~(\nu_{e}\leftrightarrow \bar{\nu}_{e}),\\
\sqrt{2}G_{F}(-n_{n}/2) &=& -\dot{\phi}/2
{}~~~(\nu_{\mu}\leftrightarrow \bar{\nu}_{\mu}).
\end{eqnarray}
Equations (8) and (9) represent the resonant
conditions for the channels
$\nu_{eL} \leftrightarrow \bar{\nu}_{eR}$ and
$\nu_{\mu L} \leftrightarrow \bar{\nu}_{\mu R}$, respectively.
In these transitions the helicity is flipped and
the electron or the muon lepton number is changed by two units.
Thus, we deal here with a qualitatively new phenomenon:
neutrino--antineutrino helicity flipping $\Delta L_{l} = 2$
resonance conversions.

In a medium with nonvanishing
$(n_{e} - n_{n}/2)$ and $n_{n}$ (like the Sun's
matter) conditions (8) and (9) cannot hold in a
nonrotating magnetic field. For fixed direction of the field rotation
only one of these two conditions can be fulfilled:
if, for example, $n_{e}>n_{n}/2$, for
$\dot{\phi}<0$ the resonance takes place in the electron
neutrino channel, $\nu_{eL} \leftrightarrow \bar{\nu}_{eR}$,
whereas for $\dot{\phi}>0$ it occurs in the muon
neutrino channel, $\nu_{\mu L} \leftrightarrow \bar{\nu}_{\mu R}$.

It should be emphasized that the neutrino-antineutrino
transitions are forbidden in the lowest order of the
interactions described by the effective Hamiltonian
in eq. (3): the matrix elements of $H$ which mix
neutrino and antineutrino states of the same flavour
are zero. As we have mentioned earlier, this follows from
the fact that Majorana neutrinos cannot have
nonzero diagonal magnetic moment. In Section 4 we will show  that
the mixing between neutrinos and antineutrinos
is induced in the second order of the perturbation theory
in the interactions described by $H$.

\section{Merging and Permutation of Resonances}

\indent The energies of the flavour neutrino levels, $H_{\alpha}$,
have different $\dot{\phi}$ dependence. Consequently,
the resonance densities associated with the
various neutrino transitions
are different functions of $\dot{\phi}$. Therefore
the magnetic field rotation modifies
the level crossing scheme. In particular,
one can expect such effects as reduction (or increase)
of the spatial
distance between the resonances, or merging of resonances, or
permutation of resonances, to take place.

In what follows
we shall consider the dependence
of the level crossing scheme on the value of
$\dot{\phi}$. We shall assume for simplicity
that the chemical composition of the medium
does not change appreciably along the neutrino path,
so that $r(t)\cong const$. In this case the
level energies $H_{\alpha}$
depend on one density parameter only: $N(t)$.
It can take both positive and negative values;
in particular, $N(t)$ is always positive in the Sun,
but can be negative in a strongly neutronized
medium like the one existing in the central regions
of collapsing stars.

It proves convenient to subtract the matrix
$H_{e}\cdot I$, proportional to the unit matrix $I$,
from the Hamiltonian $H$ in eq. (3). This
is equivalent to a multiplication of all wave functions
by the same phase factor
$\exp(-iH_{e}t)$ which, obviously, does
not change the neutrino transition probabilities.
The diagonal elements of the Hamiltonian
($H-H_{e}\cdot I$) read:
\EQ
\{H_{e},H_{\bar{e}},H_{\mu},H_{\bar{\mu}}\}=
\{0,\, \; -2N-\dot{\phi}, \, \; -(1+r)N+2c_{2}\delta,
\, \; -(1-r)N+2c_{2}\delta - \dot{\phi}\}.
\EN
The resonance values of the density parameter
$N$, corresponding to the different
neutrino transitions, can
be easily deduced from eq. (10). For
flavour neutrino transitions one has:
\EQ
N_{f(\bar{f})} = \pm \frac{2c_{2}\delta}{1+r},
\EN
where the plus sign corresponds to the
$\nu_{eL}\leftrightarrow\nu_{\mu L}$ channel, and the
minus sign, to the $CP$-conjugated one
$\bar{\nu}_{eR}\leftrightarrow\bar{\nu}_{\mu R}$.
The resonance values of $N$ for the
spin-flavour conversions are equal to
\EQ
N_{sf(\bar{sf})} = \frac{\pm 2c_{2}\delta - \dot{\phi}}
{1-r},
\EN
where the plus (minus) sign corresponds to the
$\nu_{eL}\leftrightarrow\bar{\nu}_{\mu R}$
($\bar{\nu}_{eR}\leftrightarrow\nu_{\mu L}$) transitions.
Finally, the resonances in the neutrino-antineutrino
transitions $\nu_{eL}\rightarrow\bar{\nu}_{eR}$ and
$\nu_{\mu L}\rightarrow\bar{\nu}_{\mu R}$ take place
respectively for
\begin{eqnarray}
N_{e \bar{e}} = - \frac{\dot{\phi}}{2},\\
N_{\mu \bar{\mu}} = \frac{\dot{\phi}}{2r}.
\end{eqnarray}

We shall assume for simplicity that
$\dot{\phi} = const$ along the neutrino trajectory.
Then the energy levels
$H_{\alpha} = H_{\alpha}(N)$ are straight lines
(see Figs. 1 and 2). After the phase transformation
of the neutrino wave functions \,
i) $H_{e}\equiv 0$ by definition, and ii) the muon
neutrino energy level $H_{\mu}$
does not depend on $\dot{\phi}$.
The antineutrino energy levels
$H_{\bar{e}}$ and $H_{\bar{\mu}}$ have the same
$\dot{\phi}$ dependence. Consequently,
when $\dot{\phi}$ varies, $H_{e}$ and
$H_{\mu}$ do not change, whereas $H_{\bar{e}}$
and $H_{\bar{\mu}}$ are shifted by the same
amount. Finally, note that
for $r < 1$ (which is realized in the Sun)
the strongest dependence on $N$ is
exhibited by $H_{\bar{e}}$, the
dependence of $H_{\mu}$ being somewhat weaker,
while that of $H_{\bar{\mu}}$ is the weakest.

We shall assume that neutrinos are propagating
in matter with density decreasing
monotonically along the neutrino trajectory.
Consider first the possibility
$c_{2}\delta > 0$. Let us recall the level crossing scheme
in the case of nonrotating magnetic field.
With $\dot{\phi} = 0$ (Fig. 1)
there are only two resonances for $N>0$:
the spin-flavour,
$\nu_{eL}\leftrightarrow\bar{\nu}_{\mu R}$,
and the flavour (i.e.
MSW) one, $\nu_{eL}\leftrightarrow\nu_{\mu L}$.
The spin-flavour resonance
is located at higher density as compared
to the density of the MSW resonance \cite{LM,AKHM3}:
\EQ
\frac{N_{sf}}{N_{f}} = \frac {1+r}{1-r} > 1.
\EN
Therefore if $\nu_{eL}$ is produced at a
density $N_{0}>N_{sf}$ and the adiabaticity
condition is fulfilled at the spin-flavour
resonance, $\nu_{eL}$ will be converted
into $\bar{\nu}_{\mu R}$ (provided the resonances
do not overlap). If $N_{0}<N_{sf}$,
or if the adiabaticity at the spin-flavour resonance
is strongly violated, $\nu_{eL}$ will be transformed
completely or partially into $\nu_{\mu L}$,
depending on the degree of adiabaticity
at the flavour resonance ($N = N_{f}$).

Suppose now that $\dot{\phi}\ne 0$. The effects
caused by $\dot{\phi}$ being nonzero depend on
the direction of the field rotation and we shall
discuss first the case $\dot{\phi}<0$.
With $|\dot{\phi}|$ increasing, the $\bar{\nu}_{\mu R}$
and $\bar{\nu}_{eR}$ energy levels become higher
and the following changes take place.

a) $0 < |\dot{\phi}| < 2c_{2}\delta$ (Fig. 2a).
A $\nu_{eL}-\bar{\nu}_{eR}$ level crossing appears
for $N>0$, so that under certain conditions (see Section 4)
$\nu_{eL}$ can undergo a resonant conversion
into $\bar{\nu}_{eR}$.

b) $2c_{2}\delta < |\dot{\phi}| < 4c_{2}\delta/(1+r)$ (Fig. 2b).
For $N>0$ a resonance appears in the
$\nu_{\mu L}-\bar{\nu}_{eR}$ channel.
This means that two different types of
spin-flavour conversion
can take place in a medium with $N>0$:
$\nu_{eL}\leftrightarrow\bar{\nu}_{\mu R}$, and
$\nu_{\mu L}\leftrightarrow\bar{\nu}_{eR}$.
Moreover, in such a medium $\nu_{eL}$
can be transformed into $\bar{\nu}_{eR}$
via two consecutive resonant conversions:
first the MSW one $\nu_{eL}\rightarrow\nu_{\mu L}$
(at $N = N_{f}$) and then the spin-flavour one
$\nu_{\mu L}\rightarrow\bar{\nu}_{eR}$
(at $N = N_{\bar{sf}}$). We shall call this
process a two-step $\nu_{eL}\rightarrow\bar{\nu}_{eR}$
resonant conversion.

c) At $\dot{\phi} = \dot{\phi}_{1m}$, where
\EQ
\dot{\phi}_{1m} \equiv - \frac{4c_{2}\delta}{1+r}
\EN
(Fig. 2c), three resonances merge in one point:
$N_{f} = N_{\bar{sf}} = N_{e\bar{e}} = 2c_{2}\delta/(1+r)$
(see eqs. (11)-(13)). In particular, the two resonances
present in the two-step
$\nu_{eL}\rightarrow\bar{\nu}_{eR}$ transition
merge and a strong $\nu_{eL}\rightarrow\bar{\nu}_{eR}$
conversion can take place in one resonant
region (see Section 5).

d) For $|\dot{\phi}| > |\dot{\phi}_{1m}|$ (Fig. 2d)
the flavour and the spin-flavour
resonances permute and if $N_{0}<N_{sf}$,
a $\nu_{eL}$ propagating
from high to low densities will cross first the
$\nu_{eL}\leftrightarrow\bar{\nu}_{eR}$ resonance and
later the $\nu_{eL}\leftrightarrow\nu_{\mu L}$
one. If the adiabaticity condition is fulfilled, $\nu_{eL}$
will undergo efficient conversion
into $\bar{\nu}_{eR}$ in the direct
$\nu_{eL}\rightarrow\bar{\nu}_{eR}$ transition
before reaching the MSW resonance
(for which the adiabaticity parameter
can be much bigger).

For large values of $|\dot{\phi}|$
the energy level $H_{\bar{\mu}}$ is far away from
the energy domain
where the $\nu_{eL}$, $\nu_{\mu L}$ and $\bar{\nu}_{\mu R}$
levels cross,
so that in the corresponding density region
the $\bar{\nu}_{\mu R}$ decouples
from the dynamics of the other three
neutrinos (see further, Section 5).

For positive values of $\dot{\phi}$ the antineutrino
energy levels $H_{\bar{e}}$ and $H_{\bar{\mu}}$
become lower as $\dot{\phi}$ increases. The following
specific features of the level crossing scheme
are worth noticing.

e) If $\dot{\phi}>0$, a resonance in the
$\nu_{\mu L}\leftrightarrow\bar{\nu}_{\mu R}$ channel
appears for $N>0$ (Fig. 2e). If the adiabaticity
condition is fulfilled for this resonance,
$\nu_{eL}$ can undergo a transition into $\nu_{\mu L}$
through the two-step resonant process of
$\nu_{eL}\rightarrow\bar{\nu}_{\mu R}$ and
$\bar{\nu}_{\mu R}\rightarrow\nu_{\mu L}$, imitating
the flavour conversion $\nu_{eL}\rightarrow\nu_{\mu L}$.

f) For $\dot{\phi} = \dot{\phi}_{2m}$, where
\EQ
\dot{\phi}_{2m} \equiv 4c_{2}\delta \frac{r}{1+r},
\EN
(Fig. 2f) the resonances in the three channels
$\nu_{eL}\leftrightarrow\nu_{\mu L}$,
$\bar{\nu}_{\mu R}\leftrightarrow\nu_{\mu L}$, and
$\bar{\nu}_{\mu R}\leftrightarrow\nu_{eL}$ merge
in the point $N = N_{f}$.

g) When $\dot{\phi}>\dot{\phi}_{2m}$, permutation
of the flavour and the spin-flavour
resonances occurs, so that now $N_{f} > N_{sf}$ in
contrast with the case of nontwisting magnetic field.
Therefore, if a $\nu_{eL}$ moves from
sufficiently high to low density regions and the
adiabaticity condition is fulfilled in the
flavour resonance, it will be transformed into
$\nu_{\mu L}$ rather than $\bar{\nu}_{\mu R}$.

h) For $\dot{\phi} > 2c_{2}\delta$ (Fig. 2h)
the spin-flavour resonance disappears if
$N > 0$ (in this case $N_{sf} < 0$). Thus,
for $N > 0$ only two resonances exist:
in the $\nu_{eL}\leftrightarrow\nu_{\mu L}$ and
$\nu_{\mu L}\leftrightarrow\bar{\nu}_{\mu R}$ transitions,
the latter occurring at larger density. For sufficiently large
positive values of $\dot{\phi}$ the
$\bar{\nu}_{eR}$ state decouples from the
rest of the system.

Let us consider next briefly the transitions
of $\bar {\nu}_{eR}$. The most interesting
effect is the $\bar{\nu}_{eR}\rightarrow\nu_{eL}$
conversion which can take place for negative values
of $\dot{\phi}$ (see Figs. 2a-2d). It can
proceed as a direct resonant conversion (Figs. 2a,b),
or as a two-step process
$\bar{\nu}_{eR}\rightarrow\nu_{\mu L}\rightarrow\nu_{eL}$
(Fig. 2d). In addition, $\bar{\nu}_{eR}\rightarrow\nu_{eL}$
conversion is possible in the case of
merging of three resonances (Fig. 2c).
This may have important implications for
the neutrinos emitted in collapsing stars.

Finally, let us discuss the
interesting special case of $r = 1$. As it follows
from eq. (2), it corresponds to propagation of
neutrinos in an isotopically neutral medium,
$n_{e} = n_{p} = n_{n}$, and can be realized in
wide density regions in the supernovae.
For $r = 1$ the matter density
term ($N$) no longer enters
the resonant conditions for the
spin-flavour conversions
(eqs. (6) and (7)), which take
the form:
\EQ
\dot{\phi}= \pm 2c_{2}\delta.
\EN
Thus, the spin-flavour conversions
in the case under discussion can
take place only in rotating magnetic fields.
Further, for $r = 1$ the dependences
on $N$ of the $\bar{\nu}_{eR}$ and
$\nu_{\mu L}$ energy levels,
as well as those of the $\nu_{eL}$ and
$\bar{\nu}_{\mu R}$ energy levels,
are the same: the energy levels $H_{\alpha}$
as functions of $N$ form two
pairs of parallel lines (Fig. 3). For
$\dot{\phi} = - \dot{\phi}_{m} = - 2c_{2}\delta$,
merging of two pairs of resonances, namely,
of those in the channels
$\nu_{eL}\leftrightarrow\nu_{\mu L}$ and
$\nu_{eL}\leftrightarrow\bar{\nu}_{eR}$,
and in the channels
$\bar{\nu}_{\mu R}\leftrightarrow\nu_{\mu L}$
and $\bar{\nu}_{eR}\leftrightarrow\bar{\nu}_{\mu R}$,
takes place. Moreover, the levels
of $\nu_{\mu L}$ and $\bar{\nu}_{eR}$
coincide now. This means that the
$\bar{\nu}_{eR}\leftrightarrow\nu_{\mu L}$
precession will proceed with practically
maximal amplitude
for any value of $N$, except in the vicinity of
the triple resonance points.
For $\dot{\phi} = + 2c_{2}\delta$
the levels of $\bar{\nu}_{\mu R}$ and
$\nu_{eL}$ merge, and now the
$\nu_{eL}\leftrightarrow\bar{\nu}_{\mu R}$
precession will proceed with maximal amplitude.

One can perform similar analysis
in the case $c_{2}\delta < 0$. The results
can be obtained formally from those
derived above for $c_{2}\delta >0$
(see a) - h), etc.) by replacing
$c_{2}\delta$ by $-c_{2}\delta$, and by
interchanging $\nu_{lL}$ and $\bar{\nu}_{lR}$.
For instance, if $\dot{\phi}<0$ there will be again
a resonance in the transition
$\nu_{eL}\leftrightarrow\bar{\nu}_{eR}$,
and for $\dot{\phi} = \dot{\phi}_{1m} =
4c_{2}\delta/(1+r)$ three resonances,
this time in the channels
$\bar{\nu}_{eR}\leftrightarrow\bar{\nu}_{\mu R}$,
$\nu_{eL}\leftrightarrow\bar{\nu}_{eR}$
and $\nu_{eL}\leftrightarrow\bar{\nu}_{\mu R}$,
will merge in one point ($N = N_{\bar{f}}
= N_{e\bar{e}} = N_{sf} = - 2c_{2}\delta$/(1+r)).

As we see, an electron neutrino
propagating in matter and magnetic field
can be transformed into any other state of the
neutrino system. The neutrino final state
will represent, in general,
a mixture of the states of the four neutrinos:
$\nu_{eL}$, $\bar{\nu}_{eR}$, $\nu_{\mu L}$,
and $\bar{\nu}_{\mu R}$.
Their relative admixtures in the final state are determined
by i) the density $N_{0}$ at which
the neutrinos are produced, ii) the order
of the different resonances crossed by the
neutrinos, iii) the degree of overlapping
of different resonances,
and iv) by the degree to which
the adiabaticity condition is fulfilled
in each of the resonances. Magnetic field rotation
can change the order of the resonances
and also can indirectly influence
the relevant adiabaticity conditions.
In the case of nonuniform field rotation
$\ddot{\phi}$ enters the adiabaticity
conditions directly \cite{Sm,AKS}. Magnetic
field rotation can change significantly
the composition of the final neutrino state.
The most interesting new effects
from a phenomenological
point of view are the
$\nu_{eL}\leftrightarrow\bar{\nu}_{eR}$ transitions
and the merging of different resonances; we will discuss these effects
quantitatively in the next two sections.

\section{$\nu - \bar {\nu}$ Resonances}

\indent Let us consider the $\nu_{eL}\leftrightarrow\bar{\nu}_{eR}$
transitions in the vicinity of the $\nu_{eL}- \bar{\nu}_{eR}$
level crossing. As we have mentioned earlier, to lowest
order of the perturbation theory in the interactions
described by $H$ there is no
$\nu_{eL}-\bar{\nu}_{eR}$ mixing.
The latter arises as a result of the interplay of the two types
of mixing present in $H$ in eq. (3): the flavour
mixing and the mixing generated by the magnetic moment
interaction with the magnetic field. More specifically,
according to eq. (3) the transition
$\nu_{eL} \rightarrow \bar{\nu}_{eR}$ can proceed in two
ways:
\begin{eqnarray}
\nu_{eL} \longrightarrow \nu_{\mu L} \longrightarrow \bar{\nu}_{eR},\\
\nu_{eL} \longrightarrow \bar{\nu}_{\mu R} \longrightarrow \bar{\nu}_{eR},
\end{eqnarray}
i.e. via $\nu_{\mu L}$ or $\bar{\nu}_{\mu R}$ in the intermediate state.
In (19) the first transition is due to the flavour mixing
($s_{2}\delta$) and the second is induced by the interaction with
the magnetic field ($\mu B_{\bot}$), and vice versa in (20).
Since the system of evolution equations (3) formally
coincides with the Schroedinger equation, one can make use of the
perturbation theory to get the following amplitudes of the processes (19)
and (20):
\begin{eqnarray}
M_{1} &=& - s_{2}\delta\, \mu B_{\bot}\, P(\nu_{\mu L}),\\
M_{2} &=& \mu B_{\bot}\, s_{2}\delta\, P(\bar{\nu}_{\mu R}).
\end{eqnarray}
Here $P(\nu_{\mu L})$ and $P(\bar{\nu}_{\mu R})$ are the propagators of the
system with $\nu_{\mu L}$ ($\bar{\nu}_{\mu R}$) in the intermediate state:
\EQ
P(\nu_{\mu L}(\bar{\nu}_{\mu R}))= (H_{e}-H_{\mu(\bar{\mu})})^{-1}.
\EN
The opposite signs in eqs. (21) and (22) are related to the fact that
the matrix of Majorana magnetic moments of neutrinos is antisymmetric.
Evidently, for
$P(\nu_{\mu L}) = P(\bar{\nu}_{\mu R})$ a complete compensation
between the two matrix elements takes place and the
$\nu_{eL} \rightarrow \bar{\nu}_{eR}$ transition is not
possible even in the second order of the perturbation theory.
Since the self-energies of $\nu_{\mu L}$ and $\bar{\nu}_{\mu R}$
are different in matter, the cancellation between the two amplitudes
(21) and (22) is not complete.
Direct calculation in the second order of perturbation theory yields the
following amplitude of the $\nu_{eL}\rightarrow \bar{\nu}_{eR}$ transition:
\EQ
M(\nu_{eL}\rightarrow \bar{\nu}_{eR})=M_1+M_2=\frac{2s_2\delta\,\mu B_{\bot}
(rN-\dot{\phi}/2)}{[2c_2\delta-(1+r)N][2c_2\delta-\dot{\phi}-(1-r)N]}
\EN
Eq. (24) defines an effective coupling which mixes
$\nu_{eL}$ and $\bar {\nu}_{eR}$, or in other words, a
nondiagonal $\nu_{eL} - \bar {\nu}_{eR}$ term in the higher
order effective Hamiltonian of the neutrino system.

{}From eq. (24) there follow
several qualitative conclusions.
i) The $\nu_{eL} - \bar {\nu}_{eR}$ mixing can exist even in
the absence of magnetic field rotation \cite{LM,AKHM3}.
ii) Since $rN \sim n_{n}$, in a nonrotating field the
$\nu_{eL} - \bar {\nu}_{eR}$ mixing vanishes when the neutron
concentration is zero \cite{AKHM3}.
iii) For sufficiently small $n_{n}$,
the $\nu_{eL} - \bar {\nu}_{eR}$ mixing is induced by the
magnetic field rotation.
iv) The $\nu_{eL} - \bar{\nu}_{eR}$ mixing is a function of
$\dot{\phi}$, and depending on the sign of $\dot{\phi}$ (i.e. on
the direction of the field rotation) a twisting magnetic field
can either increase or reduce the $\nu_{eL} - \bar{\nu}_{eR}$ mixing,
and consequently can enhance or suppress the probability
of the $\nu_{eL} \rightarrow \bar {\nu}_{eR}$ transition.
In particular, in the $\nu_{\mu L}-\bar {\nu}_{\mu R}$
resonance point, where $\dot{\phi} = 2rN$,
one has $M = 0$.

In order to study the properties of the
$\nu_{eL} \rightarrow \bar {\nu}_{eR}$ resonant
conversion one has to solve, in general, the
system of neutrino evolution equations (3). However, in
some physically interesting cases the
$\nu_{eL} - \bar {\nu}_{eR}$ system decouples from
$\nu_{\mu L}$ and $\bar {\nu}_{\mu R}$ and its evolution
can be analyzed separately. This is realized when
the $\nu_{eL}\leftrightarrow\bar{\nu}_{eR}$ conversion
resonance is sufficiently
well separated both in density and in energy
from the resonances associated with the
$\nu_{eL}\leftrightarrow\nu_{\mu L}$ and
$\nu_{eL}\leftrightarrow\bar{\nu}_{\mu R}$ transitions.
In other words, the $\nu_{eL}-\bar{\nu}_{eR}$
resonance should not overlap with the
$\nu_{eL}-\nu_{\mu L}$ and
$\nu_{eL}-\bar{\nu}_{\mu R}$ resonances,
and the mixing of $\nu_{eL}$ and $\bar{\nu}_{eR}$
with $\nu_{\mu L}$ and $\bar{\nu}_{\mu R}$ should be
sufficiently small.

There are altogether four general conditions
ensuring the decoupling of the $\nu_{eL}-
\bar{\nu}_{eR}$ system from $\nu_{\mu L}$ and
$\bar{\nu}_{\mu R}$:
$|H_{\mu}-H_{e}| \gg 2|H_{\mu e}|$,
$|H_{\bar{\mu}}-H_{e}| \gg 2|H_{\bar{\mu} e}|$,
$|H_{\mu}-H_{\bar{e}}| \gg 2|H_{\mu \bar{e}}|$,
and $|H_{\bar{\mu}}-H_{\bar{e}}| \gg 2|H_{\bar{\mu} \bar{e}}|$.
Using the effective Hamiltonian ($H - H_{e} \cdot I$), one obtains
in the "vicinity" of the $\nu_{eL}-\bar{\nu}_{eR}$
resonance, where $|c_{2}\delta|\gg |N+\dot{\phi}/2|$,
\EQ
|H_{\mu}|, |H_{\bar{\mu}}| \gg |2s_{2}\delta|,
                   2\mu B_{\bot}.
\EN
\indent The conditions for nonoverlapping of the resonances
in density scale can be written as
\begin{eqnarray}
|N_{e\bar{e}} - N_{f(\bar{f})}| \aprge 2\,\tan 2\theta_{0}
                          \,N_{f(\bar{f})},\\
|N_{e\bar{e}} - N_{sf(\bar{sf})}| \aprge 2\, (\mu B_{\bot}
      /|c_{2}\delta \pm \dot{\phi}/2|) N_{sf(\bar{sf})},
\end{eqnarray}
where $\tan 2\theta_{0}\, N_{f(\bar{f})}$ and
$(\mu B_{\bot}/|c_{2}\delta \pm \dot{\phi}/2|)\ N_{sf(\bar{sf})}$,
are essentially the
half-widths of
the MSW and the spin-flavour resonances,
the width of the
$\nu_{eL}-\bar{\nu}_{eR}$ resonance being considerably
smaller. It is easy to show that (26) and (27) are always
satisfied when inequalities (25) hold.
In a medium with $n_{e}>n_{n}$
conditions (26) and (27) can be realized
for $\dot{\phi}<0$ and, e.g. $c_{2}\delta >0$
if, for instance, $2c_{2}\delta \ll|\dot{\phi}|$, or if
$2c_{2}\delta \gg |\dot{\phi}|$. In the first case
for $r \cong 1$ we have (see Fig. 2d)
$N_{f} \ll N_{e\bar{e}} \ll N_{sf}$,
while in the second the inequalities
$N_{e\bar{e}} \ll N_{f}, N_{sf}$ (see Fig. 2a)
hold. If in the second case, e.g.,
$\Delta m^{2}$ is sufficiently large, only the resonance
condition (8) can be fulfilled, and so
only the $\nu_{eL}\rightarrow\bar{\nu}_{eR}$ conversion
can be resonantly enhanced.

If conditions (25) hold, one can block-diagonalize
the evolution matrix in eq. (3).
The resulting $2\times2$ submatrix
governing the evolution of the
$\nu_{eL}-\bar{\nu}_{eR}$ system \footnote{In fact, this matrix
describes the evolution of $\nu_{e}'$ and $\bar{\nu}_{e}'$
states obtained as a result of block-diagonalization of the initial
Hamiltonian, but they coincide with $\nu_{e}$ and $\bar{\nu}_{e}$
up to small corrections of the order of $2\sqrt{\mu^{2}B_{\bot}^{2}+
s_{2}^{2}\delta^{2}}/|-rN+\dot{\phi}/2 \pm 2c_{2}\delta|$.}
takes the form:
\EQ
\left ( \begin{array}{cc}
 0 & 2\mu B_{\bot}\,s_{2}\delta \, \beta\\
  2\mu B_{\bot}\,s_{2}\delta \,\beta &
  -2N-\dot{\phi} + 2(s_{2}^{2}\delta^{2}-\mu^{2}B_{\bot}^{2})\beta
\end{array}\right )
\EN
The nondiagonal term, $\bar{H}_{e \bar{e}}$, in the
evolution matrix (28)
\footnote {We have again
subtracted a matrix proportional to the
unit matrix in such a way that the upper diagonal
element in the $\nu_{eL}-\bar{\nu}_{eR}$
evolution matrix (28) thus derived is zero.}
coincides with the effective
$\nu_{eL}-\bar{\nu}_{eR}$ mixing parameter
obtained earlier in the second order
of the perturbation theory, eq. (24).
In the approximation used the parameter $\beta$ can be written as
\EQ
\beta\equiv \frac {-rN+\dot{\phi}/2}{[(-rN+\dot{\phi}/2)^{2}
-(2c_2\delta)^{2}]}.
\EN
The resonance condition following from (28) is
\EQ
N+\dot{\phi}/2 = (s_{2}^{2}\delta^{2}-\mu^{2}B_{\bot}^{2})\beta.
\EN
It is more accurate than the previously found condition, eq. (8),
but the correction is small. For $\mu B_{\bot}=s_{2}\delta$
the two conditions coincide.
The $\nu_{eL}-\bar{\nu}_{eR}$ mixing angle in matter,
corresponding to (28) is given by:
\EQ
\tan 2\tilde{\theta} = - \frac{2\mu B_{\bot}\,s_{2}\delta \,\beta}
{N+\dot{\phi}/2-(s_{2}^{2}\delta^{2}-\mu^{2}B_{\bot}^{2})\beta }
\EN

Let us consider now in greater detail
the nondiagonal element $\bar{H}_{e \bar{e}}$
of the $\nu_{eL}-\bar{\nu}_{eR}$ evolution
matrix (28). This element determines the
level splitting at the $\nu_{eL}-\bar{\nu}_{eR}$
resonance, as well as the degree of
adiabaticity of the $\nu_{eL}\leftrightarrow\bar{\nu}_{eR}$
transition. It can be represented as
\EQ
\bar{H}_{e \bar{e}} = - s_{2}\delta\,\mu B_{\bot}\,
  \frac{H_{\mu}-H_{\bar{\mu}}}{H_{\mu}\,
  H_{\bar{\mu}}},
\EN
and it follows from eq. (32) that
under the conditions ensuring
the validity of the evolution matrix
(28), the $\nu_{eL}-\bar{\nu}_{eR}$ mixing
is always smaller than each of the
generic first order mixings
$s_{2}\delta$ and $\mu B_{\bot}$.
If one of the two elements $H_{\mu}$ and
$H_{\bar{\mu}}$ is much bigger than the other,
$\bar{H}_{e \bar{e}}$ will be suppressed
by the smaller of the two: e.g., if
$|H_{\bar{\mu}}| \gg |H_{\mu}|$,
we have
$\bar{H}_{e \bar{e}}
\sim - \, s_{2}\delta \, \mu B_{\bot}/H_{\mu}$.
At the $\nu_{eL}\leftrightarrow\bar{\nu}_{eR}$
resonance point
(where $N = N_{e\bar{e}} = - \dot{\phi}/2)\,\,
\bar{H}_{e \bar{e}}$
can be written as:
$$ \bar{H}^{res}_{e \bar{e}} = - s_{2}\delta \, \mu B_{\bot}
 \frac{2N_{e \bar{e}}(1+r)}
   {(2c_{2}\delta)^{2}-(1+r)^{2}N_{e\bar{e}}^{2}}
 = \mp t_{2}\, \mu B_{\bot}\, \frac{N_{e\bar{e}}/N_{f(\bar{f})}}
  {1-(N_{e \bar{e}}/N_{f(\bar{f})})^{2}}= $$
\EQ
 = -s_{2}\delta \, \mu B_{\bot} \, \frac{2(1+r)/(1-r)^{2}}
    {N_{e\bar{e}}(1-N_{sf}/N_{e\bar{e}})
     (1-N_{\bar{sf}}/N_{e\bar{e}})} \, ,
\EN
where $t_{2}\equiv \tan 2\theta_{0}$, and $N_{f(\bar{f})}$
and $N_{sf(\bar{sf})}$ are
the MSW and the spin-flavour resonance densities
determined in eqs. (11) and (12).
It follows from eq. (33)
that the $\nu_{eL}-\bar{\nu}_{eR}$ mixing term
at the resonance increases when $N_{e \bar{e}}$
approaches $N_{f}$, i.e when
the $\nu_{eL}-\bar{\nu}_{eR}$ resonance approaches
the MSW resonance
\footnote {We assume here that $c_{2}\delta > 0$.}.
However, in the
approximation we use, which implies a
sufficiently good
separation between the different resonances,
$N_{e\bar{e}}$ cannot be very close to $N_{f}$. Using criteria (25)
and choosing
$N_{e\bar{e}} = N_{f} \pm 2t_{2}\, N_{f}$ in eq. (33)
one obtains for the allowed maximal
value of $\bar{H}_{e \bar{e}}$,
\EQ
\bar{H}^{res}_{e \bar{e}} \cong t_{2}\, \mu B_{\bot}
   \frac{1\pm 2t_{2}}{1-(1 \pm 2t_{2})^{2}} \, ,
\EN

\noindent and for small values of
$t_{2}$  ($2|t_{2}| \ll 1$) one finds:
$|\bar{H}^{res}_{e \bar{e}}| \cong
\frac{1}{4} \mu B_{\bot}$. Thus, the "maximal" value of
$|\bar{H}^{res}_{e \bar{e}}|$ in the case considered
is only by a factor of 4 smaller than the spin-flavour
mixing element in $H$.

{}From (33) it follows that
$|\bar{H}^{res}_{e \bar{e}}|$ increases also
when $N_{e\bar{e}}$ tends to $N_{\bar{sf}}$,
i.e. when the $\nu_{eL}-\bar{\nu}_{eR}$
resonance approaches the
$\bar{\nu}_{eR}-\nu_{\mu L}$ resonance.
The "maximal" value of
$|\bar{H}^{res}_{e \bar{e}}|$
compatible with constraint (27)
in this case is $\frac{1}{4}|s_{2}\delta|$
and is reached for $\mu B_{\bot}/(c_{2}\delta) \ll 1$.

The mixing parameter $\bar{H}^{res}_{e \bar{e}}$
enters the expression for
the adiabaticity parameter
characterizing the
$\nu_{eL}\leftrightarrow\bar{\nu}_{eR}$
transitions: $\kappa_{R} =
(2\bar{H}^{res}_{e \bar{e}})^{2}/\dot{H}_{\bar{e},res}$,
where $\dot{H}_{\bar{e},res}$ is
the derivative of $H_{\bar{e}}$ at the
resonance,
$H_{\bar{e}}$ being the
$\nu_{eL} - \bar{\nu}_{eR}$ level splitting
(see eq. (10)).
Using the lowest order
expression for $H_{\bar{e}}$, on finds:
\EQ
\kappa_{R} = t_{2}^{2}\, (\mu B_{\bot})^{2}\,
         \frac{L_{e \bar{e}}}{N_{e \bar{e}}} \,
         \frac{2 \, (N_{e \bar{e}}/N_{f})^{2}}{[1 -
          {(N_{e \bar{e}}/N_{f})}^{2}]^{2}}.
\EN
Here $L_{e \bar{e}} =
N_{e \bar{e}}/|(\dot{N} + \ddot{\phi}/2)_{res}|$,
where the value of
$(\dot{N} + \ddot{\phi}/2)$ is
taken at the resonance point.
For $\kappa_{R} \gg 1$ the
$\nu_{eL}\rightarrow\bar{\nu}_{eR}$
transition is adiabatic and
$\nu_{eL}$ can be completely converted
into $\bar {\nu}_{eR}$. Again, when
$N_{e\bar{e}}$ approaches $N_{f}$
(and/or $N_{\bar{sf}}$) the adiabaticity
of the $\nu_{eL}\rightarrow\bar{\nu}_{eR}$
transition improves.

\section{Triple Resonance. Neutrino - Antineutrino
  Oscillations.}

\indent We shall analyze here the case
when resonances merge, which is opposite to the one
discussed in Section 4. The mutual influence
of the different neutrino transitions
becomes very strong in this case.

If $\dot{\phi} = \dot{\phi}_{1m} < 0$
(see Section 3, eq. (16)),
three level crossing points, namely, the
$\nu_{eL}- \nu_{\mu L}$,
$\bar {\nu}_{eR}- \nu_{\mu L}$,
and  $\bar {\nu}_{eR}- \nu_{eL}$,
merge at $N=N_{f}$. The condition of
merging depends on the neutrino
energy $E$; for fixed $E$ the resonances
merge at definite value of $\dot{\phi}$.
If neutrinos possess
continuous energy spectrum, the merging
will occur for a
given $\dot{\phi}<0$ at definite value of
the energy, $E_{m}$:
\EQ
E_{m} = - \frac{c_{2} \, \Delta m^{2}}{(1+r)\dot{\phi}} \, .
\EN
Thus, the specific phenomena
which take place at the merging
point will have a resonance character
as functions of energy.

Suppose $\dot{\phi} = \dot{\phi}_{1m}$ and consider the dynamics of
the neutrino system in the vicinity
of the merging point which
we shall call also point of the
triple resonance. Now three neutrino
states are participating in the
transitions and
only one state, that of $\bar{\nu}_{\mu R}$,
can decouple in the triple resonance
region. One has for this state
\EQ
H_{\bar{\mu}} = - (1-r)N +
   \frac{2c_{2}\delta \, (3+r)}
   {1+r}
\EN
If $s_{2}\delta, \mu B_{\bot} \ll H_{\bar{\mu}}$
(which implies that $s_{2} \ll 1$
and $\mu B_{\bot} \ll 4c_{2}\delta$),
it is possible to perform a $3-1$
block-diagonalization of the neutrino evolution
matrix in the triple resonance region.
This leads to the following effective
Hamiltonian for the
$(\nu_{eL}, \, \bar{\nu}_{eR}, \, \nu_{\mu L} )$
system:
\EQ
H'
{}~=~ \left (
\begin{array}{ccc}
-\frac{(\mu B_{\bot})^{2}}{H_{\bar {\mu}}} & -\epsilon & s_{2}\delta \\
 -\epsilon & -2N+\frac{4c_{2}\delta}{1+r} -
              \frac{(s_{2}\delta)^{2}}{H_{\bar {\mu}}} & -\mu B_{\bot}\\
 s_{2}\delta & -\mu B_{\bot} & - N(1+r) + 2c_{2}\delta
\end{array}\right )
\EN
where
\EQ
\epsilon = \frac {(s_{2}\delta)(\mu B_{\bot})}{H_{\bar {\mu}}}
\EN
corresponds to a small direct
$\nu_{eL}-\bar {\nu}_{eR}$ mixing of
the second order, generated by the propagation only
of $\bar {\nu}_{\mu R}$. As we shall see,
the mixing induced by the
$\nu_{\mu L}$-exchange is resonantly enhanced
and turns out to be of the order of the generic
mixings. The diagonal
elements also acquire corrections due to the
propagation of $\bar{\nu}_{\mu R}$ which
are written explicitly in (38). They modify
condition (36) and the density at the
triple crossing point by terms $\sim \epsilon$.

Let us find the level splitting and the mixing
in the neutrino system described by the
Hamiltonian (38), in the merging
point. The eigenvalues of the
Hamiltonian (38) read:
\EQ
H_{1} \cong f -\frac{\epsilon'}{2},
 \, \, H_{2} \cong - f -\frac{\epsilon'}{2}, \,
  \, H_{3} \cong \epsilon'.
\EN
Here
\EQ
f = \sqrt{(\mu B_{\bot})^{2} + (s_{2}\delta)^{2}}
\EN
and
\EQ
\epsilon ' = - \epsilon \sin 2\omega ,
\EN
where
\EQ
\tan \omega = \frac{s_{2}\delta}{\mu B_{\bot}}
\EN
is the ratio of the leading
order flavour and spin-flavour
mixing terms. The parameter $f$
determines the level splitting in the leading
order (i.e. neglecting the terms
$\sim \epsilon$): $H_{1} - H_{2} = 2f$,
$H_{1} - H_{3} = H_{3} - H_{2} = f$.
Moreover, the flavour and the spin-flavour
mixing terms enter $f$ symmetrically.
(For comparison, notice that
the splitting in the
flavour and in the spin-flavour resonance
points are equal, respectively, to
$2s_{2}\delta$ and $2\mu B_{\bot}$.)
The $H_{1}-H_{2}$ splitting does not depend
on the $\bar{\nu}_{\mu R}$-level correction
$\epsilon '$, while $H_{3}$ shifts towards
$H_{1}$ or $H_{2}$ depending on the sign of
$\epsilon '$.

The orthogonal matrix diagonalizing
the Hamiltonian (38) is given to
leading (first) order correction
in $\epsilon/f$ by
\EQ
S_{m}
{}~=~ \left (
\begin{array}{ccc}
   \frac{\sin \omega}{\sqrt{2}} + \alpha
 & \frac{\sin \omega}{\sqrt{2}} - \alpha & \cos \omega \\
 -\frac {\cos \omega}{\sqrt{2}} + \gamma &
             -\frac{\cos \omega}{\sqrt{2}} - \gamma &  \sin \omega \\
\frac{1}{\sqrt{2}} - \xi & - \frac{1}{\sqrt{2}}
    - \xi &  -\frac{\epsilon}{f}\, \cos 2\omega
\end{array}\right )
\EN
where the parameters $\alpha$,
$\gamma$ and $\xi$ are all of the order
of $\epsilon/f$:
\EQ
\alpha = \frac{1}{\sqrt{2}} \, \frac{\epsilon}{f} \, \cos \omega \, (1 -
           \frac{3}{2} \sin^{2}\omega),
\EN
\EQ
\gamma = - \frac{1}{\sqrt{2}} \, \frac{\epsilon}{f}\, \sin \omega
   (1 - \frac{3}{2} \cos^{2} \omega),
\EN
\EQ
\xi = \frac{1}{4\sqrt{2}} \, \frac{\epsilon}{f} \, \sin 2\omega.
\EN

Consider now the properties of the
various neutrino transitions
which can take place
at the merging point.
If the density and the chemical composition
of the medium, and the magnetic field
(i.e. $N$, $r$, $B_{\bot}$ and $\dot{\phi}$)
do not change
along the neutrino trajectory, the
propagation of neutrinos has
a character of pure
oscillations-precessions
with constant depths and
periods.
There are altogether
three transition channels,
$\nu_{eL}\leftrightarrow\nu_{\mu L}$,
$\bar{\nu}_{eR}\leftrightarrow\nu_{\mu L}$,
$\nu_{eL}\leftrightarrow\bar {\nu}_{eR}$, and
we shall concentrate on
the qualitatively new mode
$\nu_{eL}\leftrightarrow\bar {\nu}_{eR}$.
Using (40) and (44) we find for the
relevant oscillation-precession probability
in a medium with $N$, $B_{\bot}$, and
$\dot{\phi}$ having values satisfying the
merging condition:
\EQ
P(\nu_{eL}\rightarrow\bar {\nu}_{eR}) =
 \sin^{2} 2\omega \, \sin^{4} {\frac{1}{2}ft} +
\sin^{2} 2\omega \, \sin^{2} {\frac{3}{4} \epsilon' t} \,
  \cos ft +
  d \, \sin {\epsilon' t}\, \sin {ft},
\EN
where
\EQ
\sin^{2}2\omega = \frac{4(s_{2}\delta)^{2}\, (\mu B_{\bot})^{2}}
      {[(s_{2}\delta)^{2} + (\mu B_{\bot})^{2}]^{2}}
\EN
and $d = O(\epsilon)$.
As we have already emphasized, the
$\nu_{eL}\leftrightarrow\bar {\nu}_{eR}$
oscillations-precessions can exist only in
the presence of both vacuum mixing
and magnetic moment interaction. The
first term in (48)
corresponds to the lowest order,
obtained when the term $\epsilon$ in
(38) is neglected. The depth of the oscillations,
$\sin^{2} 2\omega$, does not exhibit
any suppression related to the
higher order direct $\nu_{eL}-\bar {\nu}_{eR}$
mixing. Moreover, in the symmetric case when the
vacuum mixing is equal to that induced
by the magnetic moment interaction,
$s_{2}\delta = \mu B_{\bot}$, the
oscillation depth becomes maximal. It should
be stressed that this is possible only in
twisting magnetic fields. For
$s_{2}\delta \neq \mu B_{\bot}$ the
oscillation depth is less than unity
and it decreases when the difference
between $s_{2}\delta$ and
$\mu B_{\bot}$ increases.

The $\nu_{eL}\leftrightarrow\bar {\nu}_{eR}$
oscillation-precession length is given by
\EQ
l_{e\bar{e}} = \frac{2\pi}{f} =
    \frac{2\pi}{\sqrt{(s_{2}\delta)^{2} + (\mu B_{\bot})^{2}}}.
\EN
For $(s_{2}\delta) \ll (\mu B_{\bot})$,
this length
$l_{e\bar{e}}$ is two times
bigger than the usual precession length,
$l_{p}$: $l_{e\bar{e}} \cong 2l_{p} =
2\pi/(\mu B_{\bot})$. For
$(s_{2}\delta) \gg (\mu B_{\bot})$
it is twice as big as the
oscillation length in matter at
resonant density, $l_{osc}$:
$l_{e\bar{e}} \cong 2l_{osc} =
8\pi E/(\Delta m^{2} \, \sin 2\theta_{0})$.
In the symmetric case when
$(s_{2}\delta) = (\mu B_{\bot})$
one has: $l_{e\bar{e}} =
\sqrt{2} l_{osc} = \sqrt{2} l_{p}$.
Finally, note that
$P(\nu_{eL}\rightarrow\bar {\nu}_{eR})$
depends on the fourth power of
$\sin (\pi t/l_{e\bar{e}})$ rather
than on the second power, as in the usual case.

These features
of the $\nu_{eL}\leftrightarrow\bar {\nu}_{eR}$
oscillations are consequences of the facts that
there are three levels involved
and that the splitting between the
levels is determined by $f$.
More precisely, according to (44),
$\nu_{eL}$ and $\bar {\nu}_{eR}$ are
(to leading order)
orthogonal combinations of two states:
$\nu_{0}$ with energy
$H_{3} \cong 0$, and $\nu'$ which
itself is a maximal mixture of two states
with opposite energies $\pm f$. The amplitudes
of the $\nu_{0}\rightarrow\nu_{0}$ and
$\nu'\rightarrow\nu'$ transitions are equal,
respectively, to 1 and $\cos ft$. Their
sum in the amplitude of the
$\nu_{eL}\rightarrow\bar {\nu}_{eR}$
transition gives $(1-\cos ft) = 2\sin^{2} \frac{1}{2}ft$.
As a consequence, the
$\nu_{eL}\rightarrow\bar {\nu}_{eR}$
oscillation probability is
proportional to $\sin^{4} \frac{1}{2}ft$.

The second term in (48) is generated by
the direct $\nu_{eL}-\bar {\nu}_{eR}$
mixing in (38). It gives a long-period
modulation of the oscillation probability.
For the corresponding modulation length,
$l_{mod}$, one finds: $l_{mod} \cong
4\pi/(3\epsilon') \gg l_{e\bar{e}}$.
The first two terms in (48)
can also be written as
$\frac{1}{4}\, \sin^{2} 2\omega \,
(1+\cos^{2} ft -
2\cos ft \, \cos \frac{3}{2} \epsilon't)$,
which implies that the modulation
leads to the oscillation depth varying
between $\sin^{2} 2\omega$
and $\frac{1}{2}\, \sin^{2} 2\omega$.

Let us turn next to the neutrino transitions
in the region of the merging point
in the case of varying matter density.
The efficiencies of the resonant conversions
depend on the degree of their
adiabaticity. In what
follows we will neglect the corrections
$\sim \epsilon$ due to the $\bar{\nu}_{\mu R}$
level. This can be done, in particular,
because the time of
propagation in the resonance region
($t \sim \Delta r_{res}$) is much
smaller than the period $\sim 1/\epsilon$.
In addition, as we have established above,
the relevant level splittings are
determined by $s_{2}\delta$ and
$\mu B_{\bot}$, the terms $\sim \epsilon$
giving only small corrections.

Consider first the dependence of the
energy levels on $N$ (Fig. 4). Its
specific features are determined
essentially by the relative magnitude
of the vacuum ($s_{2}\delta$) and the
magnetic moment induced ($\mu B_{\bot}$) mixings.
In particular, a {\it critical} value of the
ratio of the two mixing elements exists,
\EQ
\left (
\frac{s_{2}\delta}{\mu B_{\bot}}
\right )_{c} =
\tan \omega_{c} \equiv \sqrt{\frac{1+r}{1-r}}.
\EN
If $\tan \omega = \tan \omega_{c}$, the
eigenvalue $H_{3}(N)$ of the Hamiltonian
(38) will coincide with the flavour
level energy $H_{\mu}(N)$ for
all values of $N$ (Fig. 4a):
\EQ
H_{3}(N) = H_{\mu}(N) = - (1+r)N + 2c_{2}\delta.
\EN
The dependence on $N$ of the other two
eigenvalues of (38)
can also be found:
\EQ
H_{1,2} =  \frac{H_{\bar{e}}}{2} \pm
          \sqrt{\frac{H_{\bar{e}}^{2}}{4} + f^{2}},
\EN
where $H_{\bar{e}}$ is defined in eq. (10).
There is a simple interpretation of
condition (51) and the critical regime of transitions
associated with it. Indeed, eq. (51) can
be rewritten as
\EQ
\frac{|H_{e} - H_{\mu}|}{|H_{\mu} - H_{\bar{e}}|} =
   \frac{(s_{2}\delta)^{2}}{(\mu B_{\bot})^{2}}
\EN
Now, the mixing, as is well known,
leads to a "repulsion" of the levels: the energy splitting between the
levels becomes bigger than that without mixing. The larger the mixing and
the smaller the splitting between the corresponding flavour levels, the
larger the repulsion effect. In our case two states influence the $H_3$
level (which coincides with that of $\nu_{\mu L}$ at high densities). Then
eq. (54) is nothing else but a balance condition ensuring that the
repulsion of the $\nu_{\mu}$ level from the
$\nu_{eL}$ and $\bar{\nu}_{eR}$ ones
compensate each other for all densities. In other words, the shift of the
$H_3$ level resulting from the flavour mixing is compensated by that
stemming from the magnetic-moment induced mixing.
Due to this balance the
$\nu_{1}$ level does not deviate from the
$\nu_{\mu L}$ one and turns
out to be a straight
line coinciding with $H_{\mu}(N)$.

In the critical case the
minimal splittings between all levels
happen in the merging point. They are determined
by eqs. (40).

The case $\tan \omega < \tan \omega_{c}$
corresponds to relatively weak vacuum
mixing. Now the levels of the
neutrinos participating in the spin-flavour
conversion, i.e. $\bar{\nu}_{eR}$ and
$\nu_{\mu L}$, repulse each other stronger
than the $\nu_{eL}$ and $\nu_{\mu L}$
levels, and $H_{3}(N)$ will not
coincide with $H_{\mu}(N)$ (Fig. 4b).
The points of minimal splitting between the
eigenvalues are shifted away
from the merging point symmetrically
to $N_{m} \pm \Delta N_{m}$. Moreover,
the splitting itself becomes smaller than $f$.
With decreasing the ratio
$(s_{2}\delta)/(\mu B_{\bot})$
the shift of the
minima from the merging point
increases approaching the maximal value
\EQ
\Delta N_{m} = \frac{\mu B_{\bot}}{\sqrt{2(1+r)}}.
\EN
The densities $N_{m} \pm \Delta N_{m}$
correspond to the points
at which
the $\nu_{eL}$-level crosses
the levels of the eigenstates of
the Hamiltonian describing the
($\bar{\nu}_{eR}, \nu_{\mu L}$) system
when $s_{2}\delta=0$ (i.e. when
the $\nu_{eL}$-level decouples). The level
splitting in these points
\EQ
\Delta H = 2\, \sqrt{\frac{2}{3+r}} \, s_{2}\delta
\EN
is determined by the (smaller)
flavour mixing.

For $\tan \omega > \tan \omega_{c}$ the vacuum
mixing dominates over the magnetic moment
mixing. The levels of the neutrinos
taking part in the spin-flavour
conversion effectively
attract each other now (Fig. 4c). Again, there are two
points of minimal splitting, and they are
shifted with respect to the merging point.
When $s_{2}\delta/(\mu B_{\bot})$
increases the shift $\Delta N_{m}$ in the
points of minimal splitting
converges to
\EQ
\Delta N_{m} = \frac{s_{2}\delta}{\sqrt{2(1-r)}},
\EN
and the splitting in these points becomes
\EQ
\Delta H = 2\, \sqrt{\frac{2}{3-r}}\, \mu B_{\bot}.
\EN
The shifts (57) can be
deduced from the densities in the crossing points
between the $\bar{\nu}_{eR}$ level and the levels corresponding
to the eigenvalues of the
($\nu_{eL}, \nu_{\mu L}$)
Hamiltonian when $\mu B_{\bot}=0$ (i.e. when the $\bar{\nu}_{eR}$ level
decouples). The splitting is
determined only by the magnetic field induced
mixing.

The corrections due to the $\bar{\nu}_{\mu R}$
state give rise to small ($\sim \epsilon$) shifts
of the levels and to a breaking of the symmetry
of the level-crossing scheme
(with respect to the reflections
$(N-N_{m})\rightarrow -(N-N_{m})$,
$H_{i}\rightarrow -H_{i})$.

\section{Resonant Conversions in the Merging Region}

Let us discuss further the resonant
transitions in the triple resonance region.
Now three neutrino levels are involved
simultaneously in the relevant transitions and the
adiabaticity conditions are more complicated
than in the two-neutrino case. One can introduce
{\it partial} adiabaticity parameters
$\kappa^{(ij)}$, $i,j=1,2,3,$ characterizing
the "jump" probabilities between given
pairs of levels, $H_{i}(N)$ and $H_{j}(N)$.
The partial adiabaticity parameter
$\kappa^{(ij)}$ can be defined
as a ratio of the spatial width of the
corresponding resonance
region, $\Delta r_{R}^{ij}$, and the
oscillation (precession) length at
the resonance, $l_{R}^{ij}$:
$\kappa^{(ij)} = \pi \Delta r_{R}^{ij}/l_{R}^{ij}$.
The oscillation/precession length $l_{R}^{ij}$ is
determined by the minimal level splitting:
$l_{R}^{ij} \cong 2\pi /\Delta H^{min}_{ij}$,
$\Delta H^{min}_{ij} = min~\Delta H_{ij}(N) =
min~(H_{i}(N)-H_{j}(N))=
H_{i}(N^{min}_{ij})-H_{j}(N^{min}_{ij})$. The width
of the resonance $\Delta r_{R}^{ij}$ can be estimated
as the distance over which the level splitting
becomes by a factor $\sqrt{2}$ bigger
than the minimal one:
\EQ
\Delta r_{R}^{ij} \cong 2 |(\dot{N})_{res}|^{-1} \, \Delta N_{ij},
\EN
where $\Delta N_{ij} = |N-N^{min}_{ij}|$ is determined by
the condition
\EQ
|\Delta H_{ij}(N^{min}_{ij} \pm \Delta N_{ij})| =
  \sqrt{2} |\Delta H^{min}_{ij}|.
\EN
This definition is equivalent to the definition
considered in Section 4 for the two-neutrino case.

An interesting feature of the
three level system under discussion is
that the $\nu_{3}$ level approaches
(as function of $N$)
the $\nu_{1}$ and $\nu_{2}$ levels
not symmetrically with respect
to the point of minimal splitting
$N_{min}$, i.e. the way the levels approach
each other when $N\rightarrow N_{min}$
depends on whether $N < N_{min}$ or
$N > N_{min}$. (This is equivalent to
having two different dependences of
$N(t)$ on $t$ for $N>N_{min}$ and
$N < N_{min}$ (see Fig. 4)). In this case
one could introduce two adiabaticity
parameters for each pair of levels,
$\kappa_{a}^{(ij)} = \kappa^{(ij)}(N>N_{min}^{ij})$, and
$\kappa_{b}^{(ij)} = \kappa^{(ij)}(N<N_{min}^{ij})$, as
well as an average adiabaticity parameter
$\bar{\kappa}^{(ij)} = (\kappa_{a}^{(ij)}
+ \kappa_{b}^{(ij)})/2$.

Consider now the resonant transitions for different values of the ratio
$(s_{2}\delta)/(\mu B_{\bot})$.

For critical values of the
parameters (Fig. 4a) the minimal
splittings for all three
types of transitions
occur at the merging point. The adiabaticity
parameter determining the jump probability
from the $\nu_{1}$ level
(which approaches $\nu_{eL}$ for
sufficiently large values of $N$)
to the $\nu_{3}$
level $(\sim \nu_{\mu L})$ is given by
\EQ
\bar{\kappa}^{(13)} = \sqrt{2} \,
      \frac{f^{2}}{1-r^{2}} \, |(\dot{N})_{res}|^{-1}
\EN
(we have used the relations
$\Delta H(N) \cong H_{e} - H_{\mu}$
for $N>N_{min}$, and $\Delta H(N) \cong
H_{\bar{e}}- H_{\mu}$ for $N<N_{min}$,
and the fact that $\Delta H_{min} = f$,
in obtaining eq. (61)). The
adiabaticity parameter for
the transition $\nu_{2}\leftrightarrow\nu_{3}$
coincides with $\bar{\kappa}^{(13)}$. One has
for the transition $\nu_{1}\leftrightarrow\nu_{2}$
\EQ
\kappa^{(12)} = 2\sqrt{2} \, f^{2} \, |(\dot{N})_{res}|^{-1}.
\EN
Note that all adiabaticity parameters are
determined by the same splitting parameter
$f$, and if $r$ is not too close to 1, are
of the same order.
If the adiabaticity conditions for the
transitions of $\nu_{1}$ are fulfilled
($\bar{\kappa}^{(12)} \aprge 1$,
$\kappa^{(13)} \aprge 1)$, and $\nu_{eL}$
is produced far from the merging
point at $N_{0}\gg N_{m}$, it will be almost
completely converted into $\bar{\nu}_{eR}$
provided $N_{0}<N_{sf}$.
If the adiabaticity is weakly broken, $\nu_{eL}$
will be transformed with
large probability into $\bar{\nu}_{eR}$,
with rather small probability into $\nu_{\mu L}$,
and with even smaller
probability will remain $\nu_{eL}$.
For $\kappa^{(13)} \sim \kappa^{(12)}$
the probabilities of the
$\nu_{eL}\rightarrow\nu_{\mu L}$ and
$\nu_{eL}\rightarrow\bar{\nu}_{eR}$
transitions will be comparable.

Suppose next that $\nu_{\mu L}$ is
produced at $N\gg N_{m}$. Then the
neutrino system will be on the $\nu_{3}$
level and in the case of adiabatic
change of density (i.e.
$\kappa^{(31)} \aprge 1$) will
end up as $\nu_{\mu L}$ at $N=0$.
However, the flavour content
of the $\nu_{3}$
state (which coincides at high densities
with $\nu_{\mu L}$) changes with the
variation of density:
\EQ
|\nu_{3}> = \{ \mu B_{\bot}\,|\nu_{eL}> + s_{2}\delta \,
            |\bar{\nu}_{eR}> -
  (1+r)(N-N_{m})\frac{\mu B_{\bot}}
  {s_{2}\delta}\, |\nu_{\mu L}> \}\, R,
\EN
where $\mu B_{\bot}$ and $s_{2}\delta$ obey
eq. (51) and $R(N)$ is a normalization factor. Note
that the relative admixtures of the
$\nu_{eL}$ and the $\bar{\nu}_{eR}$ states
are fixed, and the $\nu_{\mu L}$
admixture changes with $N$. In the merging
point the latter is zero, i.e. the $\nu_{3}$
state is composed only of $\nu_{eL}$ and
$\bar{\nu}_{eR}$.

For the noncritical values of
$s_{2}\delta/(\mu B_{\bot})$ (Figs. 4b,c)
the results in the case of adiabatic
transitions are the same
($\nu_{eL}\rightarrow\bar{\nu}_{eR}$, etc.),
but the types of the
transitions change when the adiabaticity
conditions do not hold.
Now besides the merging point there are two
other points of minimal splitting, at
$N=N_{1}$ and at $N=N_{2}$. The splittings and
the adiabaticity conditions in these two
points are determined by the smallest
mixing (flavour or magnetic moment induced)
rather than by $f$. Thus, the "jumps"
are more probable in the points
$N_{1}$ and $N_{2}$ than in the merging point
itself. Moreover, all conditions
governing the neutrino transitions
in the regions of $N_{1}$ and $N_{2}$ are
identical (up to corrections $\sim \epsilon$).
Therefore the splittings, the adiabaticity parameters,
and the amplitudes of the transitions are the same
in $N_{1}$ and $N_{2}$ (if, of course,
$|\dot{N}|$ has the same value in the two
points).

In the case of relatively
weak vacuum mixing ($s_{2}\delta < \mu B_{\bot}$)
the $\nu_{eL}$ trajectory
crosses both minimal splitting regions.
The scheme of the transitions for
$\nu_{eL}$ is similar to that in the critical
regime: if the adiabaticity is only weakly broken,
$\nu_{eL}$ will be transformed mainly into
$\bar{\nu}_{eR}$, and with a smaller
probability into $\nu_{\mu L}$
(the probability of $\nu_{eL}$
to remain $\nu_{eL}$ being strongly
suppressed). In the case of
strong adiabaticity violation in the
points $N_{1,2}$, $\nu_{eL}$
remains with large probability $\nu_{eL}$ and
only small and equal admixtures of
$\bar{\nu}_{eR}$ and $\nu_{\mu L}$
in the final state will appear.
Now the two-step process
$\nu_{eL}\rightarrow\nu_{\mu L}\rightarrow\nu_{eL}$,
whose probability depends on the adiabaticity
in the two minimal splitting points, can
be more efficient than the direct "jump"
$\nu_{eL}\rightarrow\nu_{eL}$ (from the
level $\nu_{1}$ to the level $\nu_{2}$)
which depends on the big splitting $2f$.

The neutrino transitions possess
interesting features also when the initial state
is $\nu_{\mu L}$. Its trajectory crosses the two
minimal splitting regions as well. For sufficiently
weak violation of the adiabaticity
$\nu_{\mu L}$ will be transformed into $\nu_{eL}$ and
$\bar{\nu}_{eR}$ with rather small
and comparable probabilities.
In the case of strong adiabaticity violation
$\nu_{\mu L}$ will be converted mainly into
$\bar{\nu}_{eR}$. Moreover, the stronger the
violation of the adiabaticity of the transition,
the smaller the $\nu_{\mu L}$ survival probability,
in contrast to the behaviour of the
$\nu_{\mu L}$ survival probability in the usual
two-neutrino case.

Suppose further that the vacuum mixing is relatively strong
($s_{2}\delta > \mu B_{\bot}$, Fig. 4c).
Now the conditions in the two minimal
splitting points, governing the conversions of
$\nu_{eL}$ produced at $N_{0}\gg N_{m}$,
can be different from
those in the previously considered case. Indeed,
at $N \cong N_{1}$ the splittings between
the $\nu_{1}$ level and the
other two levels turn out to be appreciable,
so that the passing through the first
minimum splitting region at
$N=N_{1}$ will not influence the $\nu_{eL}$
state considerably. Thus, neutrinos
produced as $\nu_{eL}$
will cross only one minimal splitting
region (at $N=N_{2}<N_{m}$) on their way from
high to low density domains and in the case of strong
adiabaticity violation in this region they will
be transformed into $\nu_{\mu L}$.
The behaviour of the neutrino system
in the case of $\bar{\nu}_{eR}$
in the initial state can be formally obtained
from that with $\nu_{eL}$ in the initial state
by interchanging the configurations
shown on Figs. 4b and 4c.

Let us comment finally
on the two mechanisms of
$\nu_{eL}\rightarrow\bar{\nu}_{eR}$
transitions. The
$\nu_{eL}\rightarrow\bar{\nu}_{eR}$
resonant conversion due to
the direct (second order)
$\nu_{eL}-\bar{\nu}_{eR}$ mixing
and the two-step conversion
involving two resonant transitions
$\nu_{eL}\rightarrow\nu_{\mu L}$
and $\nu_{\mu L}\rightarrow\bar{\nu}_{eR}$
coincide (are indistinguishable)
at the merging point.
One can distinguish the two
different mechanisms of
$\nu_{eL}\rightarrow\bar{\nu}_{eR}$
conversion
if the resonance regions of the
$\nu_{eL}\rightarrow\bar{\nu}_{eR}$,
$\nu_{eL}\rightarrow\nu_{\mu L}$, and
$\nu_{\mu L}\rightarrow\bar{\nu}_{eR}$
transitions are sufficiently well
separated.

The {\it direct} resonant
conversion takes place in one resonance
region, the corresponding resonance
condition does not depend (in the leading approximation)
on the neutrino
energy, and the energy dependence of the
conversion probability
follows from the energy
dependence of the relevant adiabaticity
parameter. The level splitting at the
resonance point is determined by the
product of the flavour and the
magnetic field induced mixing elements
and typically is rather small.
In contrast, in the case
of the {\it two-step} conversion two resonant
transitions take place in two spatial
regions which are separated
by macroscopic distance. A real $\nu_{\mu L}$
propagates between the two resonant points.
Both relevant
resonance conditions depend on
the neutrino energy. In each of the
two resonance points the mixing
is generic, and the splitting
is generated by first order mixing.
There are two adiabaticity parameters
(and conditions) which depend on the derivative
of $N$ in two different
regions of the neutrino trajectory.

Given similar conditions in the resonance
regions of the three transitions
$\nu_{eL}\rightarrow\bar{\nu}_{eR}$,
$\nu_{eL}\rightarrow\nu_{\mu L}$,
and $\nu_{\mu L}\rightarrow\bar{\nu}_{eR}$,
the two-step $\nu_{eL}\rightarrow\bar{\nu}_{eR}$
conversion will be more efficient because
the adiabaticity conditions
will be fulfilled better for each of the
two transitions involved than for the
direct conversion. Nevertheless,
under certain conditions the
direct conversion can be
the dominant
mechanism for the
$\nu_{eL}\rightarrow\bar{\nu}_{eR}$
transitions.
This can be the case in the situations corresponding to the energy level
scheme of Fig. 2d provided $N_{e\bar{e}}<N_0<N_{sf}$.
This can also happen if $N_0$ satisfies the conditions
$N_{e\bar{e}}<N_0<N_{f}$ (Fig. 2a) and if, in addition, the spin-flavour
transition $\bar{\nu}_{eR}\rightarrow \nu_{\mu L}$ is non-adiabatic
(Fig. 2b). The direct $\nu_{eL}\rightarrow \bar{\nu}_{eR}$ transitions
can also be efficient if $\bar{\nu}_{eR}$ is produced in the initial state
and propagates from $N_0>N_{e\bar{e}}$ to $N\rightarrow0$.
In all these
cases the adiabaticity condition  for the direct
$\nu_{eL}\rightarrow\bar{\nu}_{eR}$ conversion
can be fulfilled, in particular, due
to relatively small value of the
derivative $\dot{N}$.

{}From "microscopic" point of view there is
no fundamental difference between the
two mechanisms of
$\nu_{eL}\rightarrow\bar{\nu}_{eR}$
conversion discussed above. In the direct
case, the intermediate $\nu_{\mu L}(\bar{\nu}_{\mu R})$ state
is virtual and both the (first-order) flavour and spin-flavour transitions
which result in the second-order $\nu_{eL}-\bar{\nu}_{eR}$ mixing can in
principle occur at the same point. In the two-step conversion the
intermediate $\nu_{\mu L}$ state is practically real, the two resonant
conversions are separated by a macroscopic distance, and the transition
probability can be represented as a product of the individual transition
probabilities in the two resonances.
If the $\nu_{eL}\leftrightarrow\bar{\nu}_{eR}$
resonance point approaches the
$\nu_{eL}\leftrightarrow\nu_{\mu L}$  and
$\bar{\nu}_{eR}\leftrightarrow\nu_{\mu L}$
resonances, an enhancement of the
$\nu_{eL}-\bar{\nu}_{eR}$ mixing
takes place. In the merging point
the direct $\nu_{eL}-\bar{\nu}_{eR}$ mixing
is no longer suppressed and the direct
$\nu_{eL}\leftrightarrow\bar{\nu}_{eR}$
conversion becomes as efficient
as any of the transitions
caused by the generic first order mixings
present in the effective Hamiltonian
in eq. (3). At the same time when
$N_{f}$ and $N_{\bar{sf}}$ approach
$N_{m}$ the two resonant regions
associated with the two transitions
in the two-step $\nu_{eL}\rightarrow\bar{\nu}_{eR}$
conversion approach each other and
merge; both resonances take place
in the same point, thus recovering the case
of the direct $\nu_{eL}\rightarrow\bar{\nu}_{eR}$
conversion.

We have studied so far the merging of three
resonances when
$\dot{\phi} = - 4c_{2}\delta/(1+r) < 0$.
The case of merging of the
$\nu_{\mu L}\leftrightarrow\nu_{eL}$,
$\nu_{eL}\leftrightarrow\bar{\nu}_{\mu R}$,
and $\nu_{\mu L}\leftrightarrow\bar{\nu}_{\mu R}$
resonances (Fig.~2f), which can be realized for
$\dot{\phi} = 4c_{2}\delta \, r/(1+r) > 0$,
can be analyzed along the same lines.

\section{Phenomenological Implications}

\indent We would like to make a few remarks
concerning the possible phenomenological
implications of the results obtained above.
Consider first the consequences for the solar neutrinos.

1. It should be noted that magnetic
field configurations in the Sun implying
twisting fields were considered in
the astrophysical literature without
any reference to the solar neutrino problem
\cite{ASTRO}. What is really to be
questioned is whether one should expect
the magnitude of the field rotation
velocity to be relevant for the
evolution of the neutrino system.
Let us show that this is really
the case. As we have seen, the field
rotation effects are important when
\EQ
\dot{\phi}\sim \sqrt{2}G_{F}N_{eff}.
\EN
Consider neutrino spin or spin-flavour
precession in the
convective zone (precession in the interior
regions of the Sun is unable to account
for the time variations of the
solar neutrino flux).
The magnitude of $\dot{\phi}$
can be characterized by the curvature radius
$r_0$ of the magnetic field
lines, $\dot{\phi}\sim 1/r_0$. The depth of the
solar convective zone is~
$\sim 0.3 R_{\odot}$, so that one can
expect $r_0$ to be of the order of 10\%
of the solar radius $R_{\odot}$. This gives
$\dot{\phi}\sim 10/R_{\odot}
\simeq 3\times 10^{-15}$ eV, whereas the r.h.s.
of eq. (64) for the density
near the bottom of the convective zone
$\rho\simeq 0.16$ g/cc is $\simeq 8
\times 10^{-15}$ eV, i.e. exactly of the same
order of magnitude. It is
remarkable that such apparently unrelated
quantities as the Fermi constant,
solar matter density in the convective zone and
solar radius turn out to
satisfy eq. (64). We would like to emphasize
that one does not need any fine
tuning of the parameter $\dot{\phi}$ : for each
value of $\dot{\phi}$ in the
above-defined range and given neutrino
energy $E$ there exists a
corresponding value of matter density in
the convective zone for which the
resonance condition is satisfied.

2. As was pointed out in \cite{Sm,AKS}, field
rotation effects can either
enhance or suppress neutrino spin or
spin-flavour precession in the Sun
depending on the sign and magnitude of
$\dot{\phi}$. It can also change
the energy dependence of the $\nu_{eL}$
survival probability and so
distort the solar neutrino spectrum.
Twisting
magnetic field can give rise to some
specific effects which could, in principle,
be used for identification of the rotation.
For example, the picture of
the semiannual variations of the
solar neutrino flux should crucially
depend on the field rotation. These
variations can be either enhanced or suppressed,
or the minima of the flux can
become asymmetric depending on the signs of
the $\dot{\phi}$ in the northern
and southern solar hemispheres \cite{AKS,KKOT}.
The field rotation effects
can also manifest themselves in additional
time dependences of the solar
neutrino flux which are not related to
the time variation of the strength of
the magnetic field (11-yr cycle). This
could come about if the field-rotation
profile in the Sun $\phi(r)$ is not constant
in time. Another possible reason for
short-time variations can be related to
the rotation of the Sun and azimuthal
dependence of the $\dot{\phi}$
profile. In this case one can
expect the neutrino flux to vary
in time with a period comparable to the
solar day ($\simeq 27$ earth days).

3. The permutation of the resonances
will lead to a change of the helicity and the
flavour content of the neutrino final state.
If the adiabaticity conditions hold,
the change of the order of the MSW and
the spin-flavour resonances will lead
to the presence of a  considerable
$\nu_{\mu L}$ component
in the neutrino flux at Earth, instead
of a $\bar{\nu}_{\mu R}$ component.
Unfortunately, this two possible
components of the neutrino flux
cannot be distinguished in the current
and next generation experiments with
solar (and supernova) neutrinos:
the neutral current interactions of
the low energy ($E \aprle 50$ MeV)
$\nu_{\mu L}$ and $\bar{\nu}_{\mu R}$
are practically the same.

4. The most interesting
results of our study are associated with the
$\nu_{eL}\rightarrow\bar{\nu}_{eR}$
transition. As we have shown,
magnetic field rotation can
modify drastically
this transition: it can strongly
enhance or suppress the
$\nu_{eL}\rightarrow\bar{\nu}_{eR}$
conversion probability and, consequently,
the $\bar{\nu}_{eR}$ component of the neutrino
flux from the Sun. In particular, the field rotation
can induce direct $\nu_{eL}\rightarrow\bar{\nu}_{eR}$
resonant conversion, make the two-step conversion
resonant, and lead to the
phenomenon of merging of resonances. In the latter cases
a strong $\nu_{eL}\rightarrow\bar{\nu}_{eR}$ conversion
is possible.
The conditions for direct
isolated $\nu_{eL}\rightarrow\bar{\nu}_{eR}$ transitions
are more restrictive than those for the usual RSFP. In particular, given
the same magnetic field and matter density distributions, the adiabaticity
parameter for the former is generally smaller. Thus, even if the RSFP
conversion is efficient in the Sun, the direct $\nu_{eL}\rightarrow
\bar{\nu}_{eR}$ transition may not be efficient. However, the adiabaticity
condition in the triple resonance can be fulfilled in the Sun even if the
transition in the isolated $\nu_{eL}\rightarrow\bar{\nu}_{eR}$  resonance
is non-adiabatic. This condition and
the relevant adiabaticity parameters
at the merging point depend on the neutrino
energy. Thus, one can expect an enhancement of the
$\bar{\nu}_{eR}$ component of the solar neutrino
flux in a definite energy range. A sizable flux of
$\bar{\nu}_{eR}$ from the Sun would be in contradiction
with the upper limit
$\Phi({\bar \nu}_{e})\aprle (5-7)\% \,\Phi(\nu_{e})$
obtained from the analyses of
the Kamiokande II and LSD data \cite{BFMM,LSD}, and so some
regions of values of the relevant parameters
can be ruled out. Note, however, that these
constraints are valid only if the $\nu_{eL}$ are produced at a density
$N_0<N_{sf}$ or if the $\nu_{eL}\rightarrow \bar{\nu}_{\mu R}$ RSFP
conversion is non-adiabatic. This
requires either small neutrino energies or large values of $\Delta m^2$;
in the first case the limit on the flux of solar ${\bar \nu}_e$ obtained
from the data of the Kamiokande and LSD experiments,
which are only sensitive to the
high energy ${}^{8}$B neutrinos, may not
be applicable. This means that the upper
bound on the flux of solar ${\bar \nu}_e$ may be used to exclude
some values of $\dot{\phi}$ only for large
enough values of $\Delta m^2$, $\Delta m^2\aprge 2\times 10^{-7}~{\rm eV}^2$.

The resonant $\nu_e\rightarrow {\bar \nu}_e$ conversion can be
compatible with the existing upper bound on the ${\bar \nu}_e$ flux from
the Sun since the transition may not be adiabatic. It is therefore
important to look for the ${\bar \nu}_e$ flux in such solar neutrino
experiments as SNO, Borexino and Super-Kamiokande which are expected to be
very sensitive to ${\bar \nu}_e$'s.

One should stress that ${\bar \nu}_e$'s
can be produced by the Sun even in the
non-resonant case when $\dot{\phi}=0$ \cite{LM,AKHM3}; however the
resonant $\nu_e\rightarrow {\bar \nu}_e$ conversion increases the ${\bar
\nu}_e$ flux significantly and also has a clear signature which allows
one to distinguish
this mechanism from the others: the flux of ${\bar \nu}_e$'s should have a
peak at the energy defined by the condition $\sin 2\omega \cong 1$
($s_{2}\delta \cong \mu B_{\bot}$),
whereas the energy spectra of ${\bar \nu}_e$'s generated by other mechanisms
(including non-resonant RSFP+neutrino oscillations) should be smooth. It is
important to notice that the position and
the width of the peak strongly depend on the magnetic field strength
in the region of the resonance; therefore the
energy spectrum of resonantly emitted solar ${\bar \nu}_e$'s should exhibit
characteristic time dependence (in particular, the energy at which the
${\bar \nu}_e$ spectrum achieves its maximum should vary in time).

Obtaining experimental information on the relative fluxes and the energy
spectra (and their time dependences) of the $\nu_{e}$, $\bar{\nu}_{e}$
and ($\nu_{\mu}+\bar{\nu}_{\mu}+\nu_{\tau}+\bar{\nu}_{\tau})$ components of
the solar neutrino flux will be crucial for testing the above results
\cite{APS2}.

Magnetic field rotation can also be important
for the transitions of the supernova neutrinos.
The field rotation in the supernovae can be
induced by differential rotation of matter.
As we have seen, it influences the order
of the resonances which in turn will
determine the fate of various neutrino
species propagating in a collapsing star.
Since neutrinos of different species
have in general different mean energies,
their transmutations can have observable
consequences \cite{MS,AB}.
This issue will be considered in
detail elsewhere.

\section{Conclusions}

For a system of flavour neutrinos
with vacuum mixing and flavour-off-diagonal magnetic moments,
the chirality dependent level shifts induced by
the magnetic field rotation
give rise to qualitatively new effects.
The energy level scheme can be changed
drastically by the magnetic field rotation.
The location of and the distances
between the resonances are changed. Two new
types of resonances appear at nonzero densities,
namely, resonances in the neutrino-antineutrino
transitions
$\nu_{eL}\leftrightarrow\bar{\nu}_{eR}$ and
$\nu_{\mu L}\leftrightarrow\bar{\nu}_{\mu R}$.
For a definite value of the magnetic
field rotation velocity $\dot{\phi}$
(or of the neutrino energy) merging
of three resonances in one point can take place.
The types of resonances that merge depend
on the sign of $\dot{\phi}$: for
$\dot{\phi} < 0$, the
$\nu_{eL}\leftrightarrow\nu_{\mu L}$,
$\bar{\nu}_{eR}\leftrightarrow\nu_{\mu L}$ and
$\nu_{eL}\leftrightarrow\bar{\nu}_{eR}$
resonances can occur at one point,
whereas for $\dot{\phi} > 0$, the
$\nu_{eL}\leftrightarrow\nu_{\mu L}$,
$\nu_{eL}\leftrightarrow\bar{\nu}_{\mu R}$ and
$\nu_{\mu L}\leftrightarrow\bar{\nu}_{\mu R}$
resonances can merge (assuming $c_{2}\delta > 0$).
For sufficiently large values of $|\dot{\phi}|$
permutation of the resonances occurs.

The resonant $\nu_{eL}\rightarrow\bar{\nu}_{eR}$
transitions can proceed in two ways:
i) direct $\nu_{eL}\rightarrow\bar{\nu}_{eR}$
resonant conversion induced by second order
$\nu_{eL}-\bar{\nu}_{eR}$ mixing;
ii) two-step $\nu_{eL}\rightarrow\bar{\nu}_{eR}$
conversion which proceeds via two consecutive
resonant conversions
$\nu_{eL}\rightarrow\nu_{\mu L}$ and
$\nu_{\mu L}\rightarrow\bar{\nu}_{eR}$
taking place in different regions
separated by a macroscopic distance.
Either of these mechanisms can dominate in the
$\nu_{eL}\rightarrow\bar{\nu}_{eR}$
transitions, depending on the values of the parameters
$c_{2}\delta$, $\dot{\phi}$, $\mu B_{\bot}$ and $N_{0}$.

In the case of merging of three
resonances the indicated two mechanisms of
$\nu_{eL}\rightarrow\bar{\nu}_{eR}$
conversion also "merge" becoming
indistinguishable. A complete
$\nu_{eL}\rightarrow\bar{\nu}_{eR}$
conversion in one resonance region is
possible in this case.

In a medium
with constant $s_{2}\delta$, $\mu B_{\bot}$ and
$\dot{\phi}$ having the values for which
the three resonances merge,
or the isolated
$\nu_{eL}\rightarrow\bar{\nu}_{eR}$
resonance occurs, the
oscillation-precession of $\nu_{eL}$ into
its $CP$-conjugate state $\bar{\nu}_{eR}$
can proceed with maximal depth.

The effects discussed above can exist
for solar neutrinos and for the neutrinos emitted
by collapsing stars.

\section*{Acknowledgements}
The authors are grateful to M. Moretti for discussions at the early
stage of the present study. A.Yu.S. would like to thank Prof. A. Salam,
the International Atomic Energy Agency and UNESCO for hospitality at
the International Centre for Theoretical Physics.
The work of S.T.P. was supported in part by
the Bulgarian National Science Foundation via grant PH-16.

\newpage

{\large {\bf Figure Captions}}

Figure 1. Neutrino energy levels versus density $N$
   for $\dot{\phi}=0$ and $c_{2}\delta> 0 $. The $\nu_{eL}$
   and $\nu_{\mu L}$ levels
   are shown as solid lines, the
   $\bar{\nu}_{\mu R}$ and the $\bar{\nu}_{eR}$ levels
   as dash-dotted and dashed lines, respectively.
Dotted lines represent the energy levels of the eigenstates
of the Hamiltonian.

Figure 2. Neutrino energy levels versus $N$
   for different values of the magnetic field
   rotation velocity $\dot{\phi}$.
   The notations are
   the same as in Fig. 1. Figures $a-d$ correspond
   to the case of $\dot{\phi}<0$, and figures $e-h$
   correspond to $\dot{\phi}>0$;
   the consecutive figures $a-d$ and $e-h$ have been
   obtained by
   increasing the value of $|\dot{\phi}|$ (see the text for details).
   The $\nu_{eL}\rightarrow\bar{\nu}_{\mu R}$
   resonance is not shown in fugures $a-d$ because
   of lack of space.

Figure 3. The level-crossing scheme for isotopically
   neutral medium (r=1). The notations are the same as in
   Fig. 1.

Figure 4. The dependence of the neutrino energy
   levels on density in the case of triple
   resonance merging ($\dot{\phi}= - 4c_{2}\delta/(1+r) < 0$)
   for: a) critical value of the
   ratio $s_{2}\delta/(\mu B_{\bot})$ (see eq. (51)), b) weak
   flavour mixing, $\tan \omega < \tan \omega_{c}$,
   and c) strong flavour mixing, $\tan \omega > \tan \omega_{c}$.
The notations are the same as in Fig. 1.

\newpage

\begin{thebibliography}{99}
\bibitem{C} A. Cisneros, Astrophys. Space Sci. 10 (1970) 87.
\bibitem{Shrock} K. Fujikawa, R.E. Shrock, Phys. Rev. Lett. 45 (1980) 963.
\bibitem{VVO} M.B. Voloshin, M.I. Vysotsky, Sov. J. Nucl. Phys. 44
(1986) 845; M.B. Voloshin, M.I. Vysotsky, L.B. Okun, Sov. Phys.
JETP 64 (1986) 446.
\bibitem{ShV} J. Schechter, J.W.F. Valle, Phys. Rev. D24 (1981) 1883;
Phys. Rev. D25 (1982) 283 (E).
\bibitem{AKHM1} E.Kh. Akhmedov, Sov. J. Nucl. Phys. 48 (1988) 382;
Phys. Lett. B213 (1988) 64.
\bibitem{LM} C.-S. Lim, W.J. Marciano, Phys. Rev. D37 (1988) 1368.
\bibitem{MS} S.P. Mikheyev, A.Yu. Smirnov, Sov. J. Nucl. Phys. 42
(1985) 913; Prog. Part. Nuc. Phys. 23 (1989) 41.
\bibitem{W} L. Wolfenstein, Phys. Rev. D17 (1978) 2369.
\bibitem{AKHM4} E.Kh. Akhmedov, Phys. Lett. B257 (1991) 163.
\bibitem{VW} J. Vidal, J. Wudka, Phys. Lett. B249 (1990) 473.
\bibitem{ASh1} C. Aneziris, J. Schechter, Int. J. Mod. Phys.
A6 (1991) 2375.
\bibitem{Sm} A.Yu. Smirnov, Phys. Lett. B260 (1991) 161.
\bibitem{AKS} E.Kh. Akhmedov, P.I. Krastev, A.Yu. Smirnov, Z. Phys.
C52 (1991) 701.
\bibitem{ASh2} C. Aneziris, J. Schechter, Phys. Rev. D45 (1992) 1053.
\bibitem{AKHM3} E.Kh. Akhmedov, Sov. Phys. JETP 68 (1989) 690.
\bibitem{ASTRO} Y. Nakagawa, In {\em Solar Magnetic Fields}, ed. by
R. Howard, Springer-Verlag, New York, 1971.
\bibitem{KKOT} T. Kubota, T. Kurimoto, M. Ogura, E. Takasugi, Phys.
Lett. B 292 (1992) 195.
\bibitem{BFMM} R. Barbieri, G. Fiorentini, G. Mezzorani, M. Moretti,
Phys. Lett. B259 (1991) 119.
\bibitem{LSD} LSD Collaboration, M. Aglietta {\em et al.}, preprint
ICGF 269/92.
\bibitem{APS2} E.Kh. Akhmedov, S.T. Petcov, A.Yu. Smirnov, in preparation.
\bibitem{AB} E.Kh. Akhmedov, Z.G. Berezhiani, Nucl. Phys. B373 (1992) 479.

\end{thebibliography}
\end{document}